# Bi$_{1-x}$R$_x$FeO$_3$ (R=rare earth): a family of novel magnetoelectrics

Z.V. Gabbasova, M.D. Kuz'min, A.K. Zvezdin

*General Physics Institute of the USSR Academy of Sciences, 38 Vavilov Street, 117942 Moscow, USSR*

I.S. Dubenko, V.A. Murashov, D.N. Rakov

*Moscow Institute of Radio Engineering, Electronics and Automation, 78 Vernadsky Avenue, 117454 Moscow, USSR*

and

I.B. Krynetsky

*The Problem Laboratory for Magnetism, Moscow State University, 119899 Moscow, USSR*

Crystals of solid solutions Bi$_{1-x}$R$_x$FeO$_3$, where R=La, Dy, Gd, were obtained with $x \leq 0.7$. Solid solutions of the stated rare earths, as $x$ is increased from 0 to 0.7, have one and the same sequence of five crystal structures (rhombohedral C$_{3v}^6$, triclinic C$_1^1$, orthorhombic D$_2^6$, orthorhombic D$_2^5$, orthorhombic C$_{2v}^9$). The ferroelectric–paraelectric transition occurs in rhombohedral and triclinic crystals at $T=810$–$560°$C. The high temperature modifications are orthorhombic and cubic. The orthorhombic structure C$_{2v}^9$ holds up to $1180°$C. The ferroelectric domain structure was distinguished in all types of crystals. No magnetoelectric effect (MEE) was detected in the orthorhombic crystals with the D$_2$ (222) symmetry class. But the mm2 crystals were found to have both quadratic and linear MEE. The value of the quadratic effect is considerably smaller than that of the linear one. Magnetoelectric hysteresis takes place in the crystals. The tensorial properties of the obtained crystals are analyzed from the viewpoint of crystal symmetry.

## 1. Introduction

Magnetoelectrics, i.e. substances where electric and magnetic orders coexist, have been for a long time attracting explorers with their unusual properties. The interaction of electric and magnetic subsystems of a crystal, being of fundamental interest by itself, is as well promising from the practical point of view enabling one to affect the magnetic properties by applying an electric field and vice versa.

Unfortunately, few magnetoelectric substances are known thus far. One of them is bismuth ferrite BiFeO$_3$ - a compound with the rhombohedrally distorted perovskite structure. This compound is interesting by the fact that side by side with the spontaneous polarization it must also have spontaneous magnetization, since weak ferromagnetism is allowed by the symmetry. The latter fact is a rarity in magnetoelectrics (which in turn are very rare among solids). Unfortunately, in pure BiFeO$_3$ this possibility is not realized, since the spiral magnetic structure existing in it [1] leads to vanishing of the overall magnetization [2].

Thus it seems promising to suppress the spatial non-uniformity of magnetic structure in BiFeO$_3$ by substitution of rare earths for bismuth. As is well known, the compounds of RFeO$_3$, the rare earth orthoferrites, have also the perovskite structure, but distorted orthorhombically (space group D$_{2h}$-Pbnm [*1] [3]), all of them being spatially uniform weak ferromagnets [4]. In this connection it seems likely that in the Bi$_{1-x}$R$_x$FeO$_3$ system structure phase transitions must occur as $x$ varies. Besides that, sub-

---

[*1] In conformity with the long-standing tradition founded by Geller [3], a non-standard crystal axes setting is used herein.

stitution of rare earths results in not only a significant "gain" in magnetic properties at low temperatures, but as well makes them strongly anisotropic, which might lead, like in orthoferrites, to various spin-reorientation transitions [4].

We have realized the stated substitution and obtained the family of $Bi_{1-x}R_xFeO_3$ single crystals possessing, at R=Dy within a certain range of concentration $x$, record magnetoelectric parameters. This Letter presents the experimental and theoretical study of these compounds.

## 2. Sample preparation

We obtained single crystals of both "pure" $BiFeO_3$ and solid solutions on its basis. Under the usual spontaneous crystallization in the $Bi_2O_3$–$Fe_2O_3$ system, only aggregations of small (up to 1 mm) dendrite crystals of $BiFeO_3$ were obtained, a result similar to that of ref. [5]. Structurization of melt prevented growing bigger crystals. Various additions were tried to suppress structurization. The best result was achieved with an addition of 6.5% of NaCl. Thus the optimal composition of the charge was as follows: 75.6 mole% $Bi_2O_3$, 17.9 mole% $Fe_2O_3$ and 6.5 mole% NaCl. By spontaneous crystallization of this melt (with a cooling rate of 0.5°C/h within the interval 870–820°C) the $BiFeO_3$ crystals were obtained with the pseudo-cubic property and size up to 10 mm on edge. The growth was conducted in a thin (up to 10 mm) layer of melt with reversible crucible rotation; on growth termination ($T=820°C$) the crystals were extracted from the melt by a platinum net.

Similarly, in the $Bi_2O_3$–$Fe_2O_3$–$R_2O_3$–NaCl system (R=La, Sm, Gd, Dy) the crystals of the solid solutions $Bi_{1-x}R_xFeO_3$ were obtained with $x \leqslant 0.7$. The specific feature of the process in this case consisted in the increase of the crystallization temperature (up to 1170°C for $Bi_{0.3}La_{0.7}FeO_3$) and a high value of the crystal-melt distribution factor $K_{R_2O_3}$ (running at 15 for $Bi_{0.5}Dy_{0.5}FeO_3$). The crystal composition change during the crystallization ($\Delta x \pm 0.03$) was the consequence of such high $K$'s. The conditions of crystal growth for some compositions are given in table 1.

The crystals of solid solutions had the pseudo-cu-bic property as well: {100} facets for the rhombohedral and triclinic crystals and (001, 110, 1$\bar{1}$0) for the orthorhombic ones. The typical size was ~4 mm on edge.

Table 1
Crystallization conditions and crystal compositions for $Bi_{1-x}R_xFeO_3$. In each case the charge contained an addition of 6.5 mole% of NaCl.

| No. | Charge composition (mole%) | | Crystallization interval $T$ (°C) | Crystal composition $x$ |
|---|---|---|---|---|
| 1 | $Bi_2O_3$ | 74.8 | 940–860 | 0.21 |
|   | $Fe_2O_3$ | 17.75 | | |
|   | $La_2O_3$ | 0.95 | | |
| 2 | $Bi_2O_3$ | 68.7 | 1170–1100 | 0.70 |
|   | $Fe_2O_3$ | 20.65 | | |
|   | $La_2O_3$ | 3.35 | | |
| 3 | $Bi_2O_3$ | 73.0 | 1010–975 | 0.45 |
|   | $Fe_2O_3$ | 18.0 | | |
|   | $Sm_2O_3$ | 2.5 | | |
| 4 | $Bi_2O_3$ | 75.8 | 950–920 | 0.30 |
|   | $Fe_2O_3$ | 16.5 | | |
|   | $Gd_2O_3$ | 1.2 | | |
| 5 | $Bi_2O_3$ | 73.5 | 970–920 | 0.55 |
|   | $Fe_2O_3$ | 17.2 | | |
|   | $Dy_2O_3$ | 2.8 | | |

Solid solutions of the stated rare earths, as $x$ is increased from 0 to 0.7, have one and the same sequence of five crystal structures. Thus, for R=La ($T=293$ K):

(1) $0 < x < 0.06$: rhombohedral R3c ($C_{3v}^6$);
(2) $0.06 < x < 0.24$: triclinic P1 ($C_1^1$);
(3) $0.24 < x < 0.40$: orthorhombic C222 ($D_2^6$);
(4) $0.40 < x < 0.55$: orthorhombic C222 ($D_2^3$);
(5) $0.55 < x < 0.70$: orthorhombic Pbn2$_1$ ($C_{2v}^9$)[#2].

With decreasing $R^{3+}$ ion radius, the concentration boundaries of the above structures in $Bi_{1-x}R_xFeO_3$ shift towards smaller $x$.

## 3. Experimental results

Samples of $Bi_{1-x}R_xFeO_3$ single crystals have been studied. Three crystal structures are ferroelectric, respectively with $C_{3v}^6$ (R3c), $C_1^1$ (P1) and $C_{2v}^9$ (Pbn2$_1$)

---

[#2] We find it convenient to use a non-standard crystal axes setting, which differs from that of ref. [6] by the permutation $a \leftrightarrow b$.

space groups which was confirmed by the presence of pyroelectricity. The vector $\gamma$ of the pyroelectric effect is oriented along the three-fold axis ([111] axis in the rhombohedral setting) and the value of the pyroelectric constant increases with increasing concentration $x$ for the rhombohedral crystals, $\gamma = [(0.4 \pm 0.05) - (0.9 \pm 0.05)] \times 10^{-6}$ C/m$^2$ K. In the triclinic crystals $\gamma = (10 \pm 1.5) \times 10^{-6}$ C/m$^2$ K and it is oriented along [011]. For the crystals with orthorhombic ($C_{2v}^9$) structure, $\gamma \parallel [001]$ and $\gamma = (0.2 \pm 0.03) \times 10^{-6}$ C/m$^2$ K.

The existence of a ferroelectric–paraelectric transition has been derived from the high-temperature RFA in the $T = 810–560°$C temperature range for rhombohedral and triclinic crystals. The high temperature modifications are orthorhombic and cubic. The orthorhombic structure $C_{2v}^9$ holds up to 1180°C.

The ferroelectric domain structure was distinguished in all types of crystals. Two types of domains appear in BiFeO$_3$ and three of these were present in the structure-5 ($C_{2v}^9$) crystals. The domains in BiFeO$_3$ were described as ferroelectric and ferroelastic ones in ref. [7]. Voltages up to 60 kV/cm applied to the crystal did not change the domain structure, like in ref. [5]. It should be noted that the measurements were performed at a low temperature ($T = 77$ K) because polarization of the crystals becomes impossible near the Curie point due to their high conductivity.

As far as the magnetic properties are concerned, BiFeO$_3$ is an antiferromagnet. The magnetization dependence remains linear in magnetic fields up to 12 MA/m and the magnetic susceptibility is equal to $\chi = 5 \times 10^{-6}$ cm$^3$/g at $T = 293$ K. In the rhombohedral crystals $\chi$ increases slightly with increasing $x$ (fig. 1a). The other four types of crystals are antiferromagnets as well, the Néel temperature $T_N$ increasing linearly as the La concentration $x$ increases (fig. 1b). The saturation magnetization in the triclinic crystals is approximately the same for all crystallographic axes, $m_0 = 0.05$ G cm$^3$/g (for R = La). The value of $\chi$ changes depending on $x$ and it has maxima near the structural phase boundaries (rhombohedral–triclinic and triclinic–orthorhombic) (fig. 1a). In the La-dope orthorhombic crystals the value of $m_0$ along the $c$-axis is 0.35–0.50 G cm$^3$/g (fig. 1a).

The magnetization has been measured on the

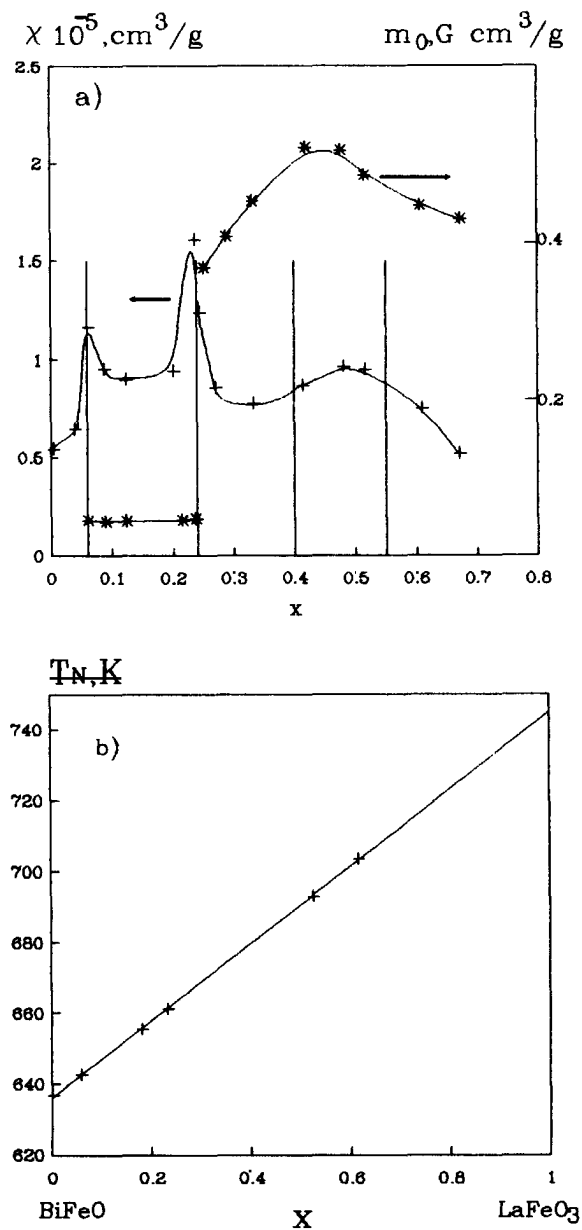

Fig. 1. Coefficient dependences of paramagnetic susceptibility $\chi$ and remnant moment $m_0$ for the Bi$_{1-x}$La$_x$FeO$_3$ system at $T = 293$ K (a) (vertical lines denote boundaries of the structure) and the Néel temperature $T = T_N$ (b).

Dy$_{0.55}$Bi$_{0.45}$FeO$_3$ single crystal along the crystal's principal axes ($a$, $b$ and $c$) in the temperature range 4.2–120 K in fields up to 60 kOe. We analyze the results by comparison with Dy orthoferrite regarded as the basic compound. Fig. 2 shows the temperature dependence of the magnetic susceptibility along the principal directions. It can be seen that, like in $DyFeO_3$ [4], the anisotropy of the magnetic susceptibility is determined mainly by $Dy^{3+}$ ions. As a distinguished feature of the mixed compound, the essential temperature dependence of $\chi_c$ should be noted ($\chi_c$ has a Van Vleck character in $DyFeO_3$, i.e. it does not depend on the temperature when $T < 70$ K). The values of magnetization in $Dy_{0.55}Bi_{0.45}FeO_3$ in strong fields ($H \sim 60$ kOe) approach the corresponding values in $DyFeO_3$ [4]. The behaviour of the spontaneous magnetization in $Dy_{0.55}Bi_{0.45}FeO_3$ in principle differs from that of $DyFeO_3$. $Dy_{0.55}Bi_{0.45}FeO_3$ appears to be characterized by $m_0$ existing along all crystallographic axes: $m_0^b = 5.5$ G cm$^3$/g, $m_0^a \sim m_0^c = 2.3$ G cm$^3$/g at $T = 4.2$ K. This result is incompatible with the symmetry and seems to be connected with the 90° ferroelectric domain structure arising as $a \sim b \sim c/\sqrt{2}$.

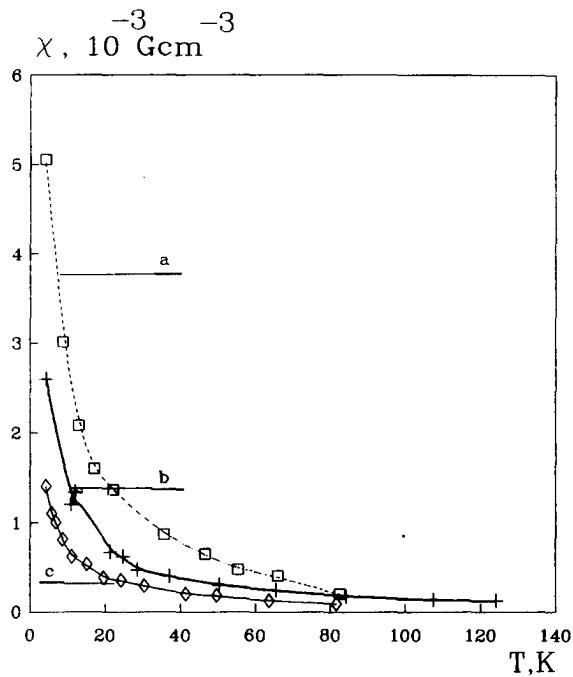

Fig. 2. Temperature dependences of $\sigma_0$ for $Dy_{0.55}Bi_{0.45}FeO_3$ single crystal obtained along three crystallographic axes.

The magnetoelectric effect (MEE) was studied by means of a static method measuring the charge arising on a capacitor containing the sample under the change of magnetic field $H$. In $BiFeO_3$ the quadratic MEE occurs at $T = 4.2$ K and 77 K. For the specimens cut parallel to the (001) and (110) planes all four components of the magnetoelectric susceptibility tensor $\beta$ were derived from the dependences of the charge on the field direction (see table 2). The values are a bit smaller than those reported in ref. [5] for $T = 4.2$ K. For $T = 77$ K the results are obtained for the first time. In triclinic crystals the effect is about 5 and 8–12 times as great as in BiFeO$_3$ at $T=4.2$ K and 77 K, respectively (see table 2). The magnitude of $\beta$ was found to be almost independent of the sort of rare earth.

Table 2
Magnetoelectric susceptibility $\beta$ of the quadratic effect in $Bi_{1-x}La_xFeO_3$ crystals.

| Crystal composition $x$ | Orientation | $\beta$ ($10^{-20}$ s/A) | |
|---|---|---|---|
| | | $T = 4.2$ K | $T = 77$ K |
| 0 | {001} cuts | 3.2 ± 0.2 | 1.7 ± 0.2 |
| | $\beta_{111}$ | 2.5 ± 0.5 | 2.0 ± 0.5 |
| | $\beta_{113}$ | −5.0 ± 0.5 | −3.6 ± 0.5 |
| | $\beta_{311}$ | −9.4 ± 0.5 | −4.0 ± 0.5 |
| | $\beta_{333}$ | 13.3 ± 0.5 | 5.3 ± 0.5 |
| 0.08 | (001) cut | 16.2 ± 0.2 | 13.3 ± 0.2 |
| 0.22 | (001) cut | | 19.7 ± 0.2 |

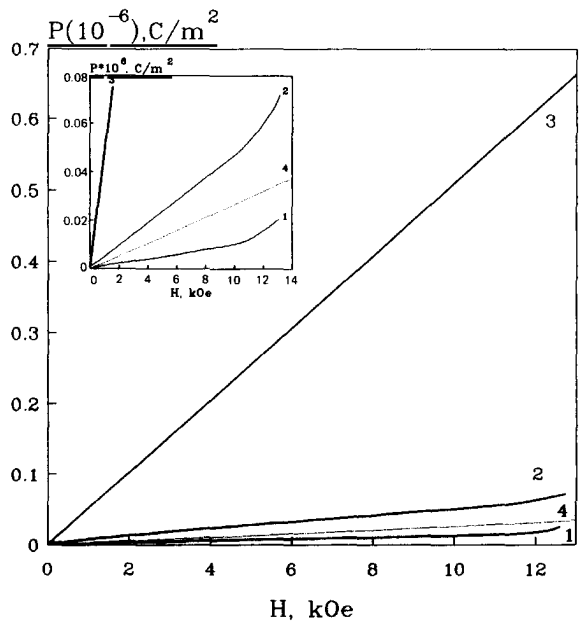

Fig. 3. Magnetoelectric effect in the orthorhombic crystals (symmetry class mm2) $Bi_{1-x}R_xFeO_3$ (plane (001), $H \perp (001)$) (1) R = La, $x = 0.55$; (2) R = Gd, $x = 0.45$; (3,4) R = Dy, $x = 0.55$; (1–3) $T = 4.2$ K; (4) $T = 77$ K.

MEE was not detected in the orthorhombic crystals with the D$_2$ (222) symmetry class. But the mm2 crystals (structure 5) were found to have both quadratic and linear MEE. The value of the quadratic effect is considerably smaller than that of the linear one. The quadratic MEE becomes noticeable for fields greater than 10 kOe (fig. 3, bends on curves 1, 2). The value of the linear MEE is strongly dependent on the sort of R-ion. It is predominant in the Dy-doped crystals, and in the crystals doped with Sm there is no MEE at all. As the temperature decreased the strength of the effect decreased rapidly and at $T=77$ K the linear effect was observed only in the Dy crystals (fig. 3, curve 4). Magnetoelectric hysteresis similar to that observed in ref. [8] takes place in the crystals. The coercive field is small and depends on the sample orientation. For the longitudinal MEE ($H, P \parallel [001]$, fig. 4) $H_c=0.55$ kOe.

At room temperature all the crystals are semiconductors (the sample resistance varied from $10^3$ to $10^5$ $\Omega$). Observation of MEE is possible only when the resistance of the sample exceeds $10^{12}$ $\Omega$, i.e. below 200 K (however, as shown by the symmetry analysis, linear MEE is expected to exist up to $T_N \sim 700$ K and quadratic MEE up to $T_C \sim 1000$ K).

## 4. Symmetry analysis

In this section some tensorial properties of the obtained crystals are analyzed from the viewpoint of crystal symmetry. Of course, the magnetoelectric properties are of principal interest herein, and, in this connection only two of the structures (Nos. 1 and 5, see section 2) are worth attention, which could exhibit both spontaneous polarization and magnetization (i.e. weak ferromagnetism). Structure 1 (pure BiFeO$_3$) has been analyzed elsewhere [2] and shown not to possess spontaneous magnetization or linear magnetoelectric effect due to its spiral magnetic structure. Thus, in this work we concentrate on structure 5, the most interesting phase as judged by its symmetry as well as by the record magnetoelectric parameters of the concerned crystal with R=Dy.

To perform the analysis it is convenient to present a polarized structure-5 crystal as a result of dipole ordering of a certain paraelectric crystal. We take as the latter the structure of a pure orthoferrite RFeO$_3$ and consider the anticipated structural phase transition D$_{2h}^{16}$ (Pbnm)–C$_{2v}^9$ (Pbn2$_1$) as the electric dipole ordering, which occurs in Bi$_{1-x}$R$_x$FeO$_3$ as $x$ is decreased from 1 to a certain critical value. All symmetry considerations will hereafter refer to the "forephase" or the "disordered phase", i.e. to the orthoferrite structure, space group D$_{2h}^{16}$. As the order parameters to describe the structural phase transition the atom displacements from their higher-symmetry positions can be used, $\zeta_R$, $\xi_{01}$, $\eta_{01}$, $\zeta_{01}$ and $\zeta_{03}$ (see table 3), which determine the z-component of the polarization. In a simple point-charge approximation one has

$$P_z = \frac{4e}{ab}(3\zeta_R - 4\zeta_{01} - 2\zeta_{03}) . \qquad (1)$$

To describe magnetic order in orthoferrites the following vector order parameters are used [4]:

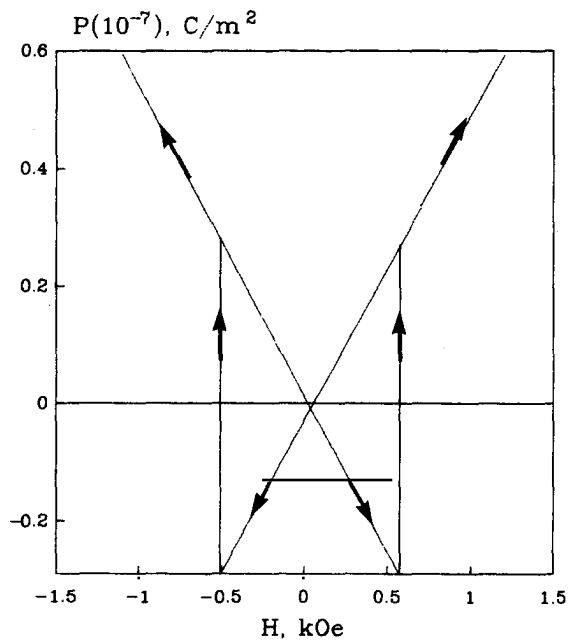

Fig. 4. Magnetoelectric hysteresis in Dy$_{0.55}$Bi$_{0.45}$FeO$_3$ crystal ((001) cut, $H \perp$ (001)).

Table 3
Atom positions for structure 5, space group $C_{2v}^9$ (Pbn2$_1$). When the displacements denoted with Greek letters become zero, the whole structure turns into that of rare earth orthoferrites, space group $D_{2h}^{16}$ (Pbnm).

| Atoms | Wyckoff symbols for $D_{2h}^{16}$ | Coordinates | | |
|---|---|---|---|---|
| | | $x$ | $y$ | $z$ |
| Bi/R | 4c | $x_R$ | $y_R$ | $\frac{1}{4}+\zeta_R$ |
| Fe | 4b | 0 | $\frac{1}{2}$ | 0 |
| O$_1$ | 8d | $x_{01}+\xi_{01}$ | $y_{01}+\eta_{01}$ | $z_{01}+\zeta_{01}$ |
| O$_2$ | 8d | $x_{01}-\xi_{01}$ | $y_{01}-\eta_{01}$ | $\frac{1}{2}-z_{01}+\zeta_{01}$ |
| O$_3$ | 4c | $x_{03}$ | $y_{03}$ | $\frac{1}{4}+\zeta_{03}$ |

$$F = V_c^{-1}(M_1 + M_2 + M_3 + M_4),$$
$$G = V_c^{-1}(M_1 - M_2 + M_3 - M_4),$$
$$A = V_c^{-1}(M_1 - M_2 - M_3 + M_4),$$
$$C = V_c^{-1}(M_1 + M_2 - M_3 - M_4), \quad (2)$$

where the $M_i$ are the mean magnetic moments of the four iron ions in a unit cell, and $V_c = abc$ is the cell volume. The vector $F$ corresponds to the magnetization; $G$, $A$ and $C$ are antiferromagnetic vectors.

The second step is to investigate the transformation properties of the introduced order parameters with respect to the symmetry operations constituting the group of the paraelectric–paramagnetic crystal. To investigate the macroscopic properties of the system it is not necessary to take into account the whole space group $D_{2h}^{16}$, but rather a reduced group $\tilde{D}_{2h}^{16}$, where all integer fundamental translations are set equal to the identity transformation. On the other hand, the time-reversal transformation must be incorporated into the group, which can be symbolically written as $\tilde{D}_{2h}^{16} = D_{2h}^{16} * C_t$, where $C_t$ denotes the second-order group comprising the time-reversal and the identity transformations. Thus it is the group $\tilde{D}_{2h}^{16}$ with respect to which transformation properties are to be studied. It is convenient to represent this group as $\tilde{D}_{2h}^{16} = C_{2v}^9 * C_i * C_t$, where $C_{2v}^9$ is the reduced group of the polarized phase.

$\tilde{D}_{2h}^{16}$ has clearly 16 one-dimensional irreducible representations (table 4), the first 8 of which are time-even and the others (primed) are time-odd. Furthermore, eight of them are spatially even (subscripts from 1 to 4) and eight are spatially odd (from 5 to 8). Thus in our approach electro-dipole and magnetic orders enjoy equal rights, and the temporal and spatial inversions are just the two operations of the symmetry group. Electric order parameters must belong to the time-even but spatially odd irreducible representations, $\Gamma_5-\Gamma_8$, those which occur in structure 5 belong to $\Gamma_8$. Magnetic order parameters must, on the other hand, belong to the time-odd but spatially even representations.

At last, we turn to our main problem, i.e. to deduce the symmetry of tensorial properties of structure-5 crystals from the crystal symmetry. We expand the polarization and the magnetization into series of powers of the electric and magnetic field components,

$$P_i = P_i^{(s)} + \kappa_{ij}E_j + \alpha_{ij}H_j + \gamma_{ijk}E_jE_k$$
$$+ \alpha_{ijk}H_jE_k + \tfrac{1}{2}\beta_{ijk}H_jH_k,$$
$$F_i = F_i^{(s)} + \chi_{ij}H_j + \alpha_{ji}E_j + \delta_{ijk}H_jH_k$$
$$+ \beta_{jik}E_jH_k + \tfrac{1}{2}\alpha_{jik}E_jE_k, \quad (3)$$

where $P^{(s)}$ and $F^{(s)}$ are the spontaneous polarization and magnetization, $\kappa_{ij} = \kappa_{ji}$ and $\chi_{ij} = \chi_{ji}$ are the electric and magnetic susceptibilities, $\alpha_{ij}$ is the linear magnetoelectric susceptibility, $\gamma_{ijk}$ and $\delta_{ijk}$ are the absolutely symmetric tensors of the non-linear electric and magnetic susceptibilities, and $\alpha_{ijk} = \alpha_{kji}$ and $\beta_{ijk} = \beta_{ikj}$ are the tensors of the quadratic magnetoelectric effect.

We state that every term on the right-hand sides of (3) must belong to that irreducible representation of $\tilde{D}_{2h}^{16}$ to which the corresponding left-hand part belongs. The field components belong clearly to the same irreducible representations as the conjugate response functions (see table 4), so that the scalar products $P \cdot E$ and $F \cdot H$ are invariants. Taking all this into account and making use of the multiplication table for the irreducible representation (see, e.g., ref. [4]) we can determine to which representation this or that tensor component belongs. The results of this procedure are also shown in table 4.

## 5. Discussion

Thus, the crystals of structure 5 can be considered as electrodipole ordered (polarized) according to the $\Gamma_8(P_z)$ mode orthoferrite. Concerning the magnetic

496

Table 4
Tensorial components of the properties under review arranged according to the irreducible representations of $\tilde{D}_{2h}^{16}$.

| | | | |
|---|---|---|---|
| $\Gamma_1$ | | $\kappa_{11}, \kappa_{22}, \kappa_{33}$ | $\chi_{11}, \chi_{22}, \chi_{33}$ |
| $\Gamma_2$ | | $\kappa_{23}$ | $\chi_{23}$ |
| $\Gamma_3$ | | $\kappa_{13}$ | $\chi_{13}$ |
| $\Gamma_4$ | | $\kappa_{12}$ | $\chi_{12}$ |
| $\Gamma_5$ | | $\gamma_{123}$ | $\beta_{123}, \beta_{213}, \beta_{312}$ |
| $\Gamma_6$ | $P_x, E_x$ | $\gamma_{111}, \gamma_{122}, \gamma_{133}$ | $\beta_{111}, \beta_{122}, \beta_{133}, \beta_{212}, \beta_{313}$ |
| $\Gamma_7$ | $P_y, E_y$ | $\gamma_{112}, \gamma_{222}, \gamma_{233}$ | $\beta_{112}, \beta_{211}, \beta_{222}, \beta_{233}, \beta_{323}$ |
| $\Gamma_8$ | $P_z, E_z, \zeta_R, \xi_{01}, \eta_{01}, \zeta_{01}, \zeta_{03}$ | $\gamma_{113}, \gamma_{223}, \gamma_{333}$ | $\beta_{113}, \beta_{311}, \beta_{223}, \beta_{322}, \beta_{333}$ |
| $\Gamma'_1$ | $G_y$ | $\delta_{123}$ | $\alpha_{123}, \alpha_{132}, \alpha_{213}$ |
| $\Gamma'_2$ | $F_x, G_z, H_x$ | $\delta_{111}, \delta_{122}, \delta_{133}$ | $\alpha_{111}, \alpha_{122}, \alpha_{133}, \alpha_{212}, \alpha_{313}$ |
| $\Gamma'_3$ | $F_y, H_y$ | $\delta_{112}, \delta_{222}, \delta_{233}$ | $\alpha_{112}, \alpha_{121}, \alpha_{222}, \alpha_{233}, \alpha_{323}$ |
| $\Gamma'_4$ | $F_z, G_x, H_z$ | $\delta_{113}, \delta_{223}, \delta_{333}$ | $\alpha_{113}, \alpha_{131}, \alpha_{223}, \alpha_{232}, \alpha_{333}$ |
| $\Gamma'_5$ | | $\alpha_{11}, \alpha_{22}, \alpha_{33}$ | |
| $\Gamma'_6$ | | $\alpha_{23}, \alpha_{32}$ | |
| $\Gamma'_7$ | | $\alpha_{13}, \alpha_{31}$ | |
| $\Gamma'_8$ | | $\alpha_{12}, \alpha_{21}$ | |

order, as was indicated above direct magnetic measurements do not enable us to determine its type. Presumably, this is connected with the 90° ferroelectric domain structure. But it is safe to assume that the crystal is ordered according to the $\Gamma'_4(F_zG_x)$ mode below $T_N$, i.e. $\boldsymbol{P} \parallel \boldsymbol{F} \parallel [001]$ within each domain. Such a magnetic structure is stabilized by the magnetic anisotropy of the Fe subsystem and is usually observed in orthoferrites with dilated R subsystem. The magnetoelectric measurements employed in this Letter give a direct confirmation of a given assumption.

Let us regard table 4. Only those tensor components must be nonzero which occupy the first line of the table ($\Gamma_1$) in the paraelectric–paramagnetic phase, i.e. they are invariants of the $D_{2h}^{16}$ group. As a result of electrodipole ordering the symmetry will reduce and tensor components belonging to $\Gamma_8$ must be added. Then, in addition those corresponding to $\Gamma'_4$ appearing due to magnetic ordering below $T_N$, become nonzero. Thus, in the polar phase the quadratic MEE $P_i = \beta_{ijk}H_jH_k/2$ is to be observed and the other quadratic MEE $F_i = \alpha_{ijk}E_jE_k/2$ below $T_N$ appears as well.

Besides that components belonging to $\Gamma'_5 = \Gamma_8 \times \Gamma'_4$ become nonzero due to interaction of the ordered electric and magnetic subsystems; the linear MEE becomes permitted and the matrix of magnetoelectric susceptibility must be diagonal in the principal axes of the rhombic crystal. It is easy to see that the ferroelectric domain structure retains the magnetoelectric susceptibility in the diagonal form. The $z$ component of the polarization measured in our experiments as a function of the field is given by the expression

$$P_z = \alpha_{33} H_z \tag{4}$$

in the first approximation.

Fig. 5 represents the orientational dependence of the MEE, i.e. the magnitude of $P_z$ plotted as a radius-vector along the field direction within the $(1\bar{1}0)$ plane, at $H = 1$ MA/m. As is seen from the figure the experimental dependence (continuous curve) is well described by eq. (4) (dashed curve) everywhere, except for the areas where the field is oriented almost perpendicular to the $z$-axis.

It is characteristic that the sign of polarization at the positive and negative magnitude of $H_z$ is positive since the sign of $\alpha_{33}$ changes with the sign changing of the magnetization $F_z$. This is clearly seen from the MEE field dependence in $Dy_{0.55}Bi_{0.45}FeO_3$ at $H \parallel [001]$ (fig. 4). That discovers the original hysteresis of the "butterfly" type when the absolute value of the magnetic field $z$ component does not exceed the coercive field $H_c$; in this case additional narrow petals where $P_z$ has negative sign appear on the MEE (fig. 5) orientational dependence in the vicinity of $H \perp [001]$.

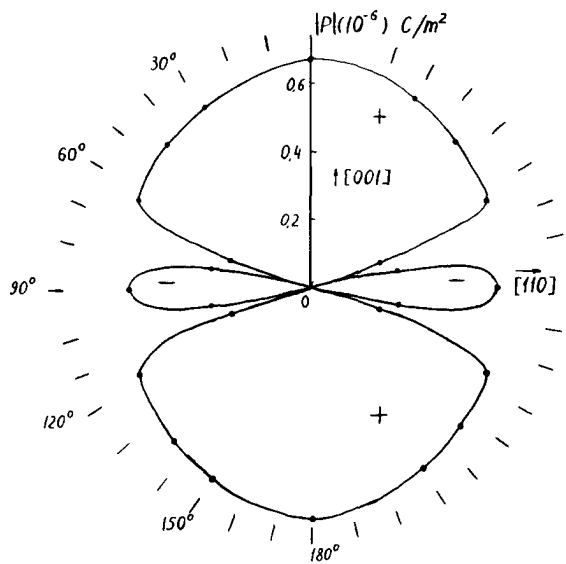

Fig. 5. z-component of the polarization of the $Dy_{0.55}Bi_{0.45}FeO_3$ single crystal induced by the magnetic field, plotted as a radius-vector along the field direction within the $(1\bar{1}0)$ plane.

In this region the sample generally breaks into magnetic domains ($\pm F_z$) and the form of narrow petals generally depends on the prehistory of the sample.

Thus, the observed MEE dependences upon the value and direction of the magnetic field in $Dy_{0.55}Bi_{0.45}FeO_3$ show the diagonal form of the MEE linear tensor in the principal axes of the rhombic crystal which in turn unequivocally shows the magnetic ordering according to the $\Gamma'_4$ state.

## References

[1] I. Sosnowska, T. Peterlin-Neumaier and E. Steichele, J. Phys. C 15 (1982) 4835.
[2] A.K. Zvezdin and M.D. Kuzmin, submitted to Physica B.
[3] S. Geller, J. Chem. Phys. 24 (1956) 1236.
[4] K.P. Belov, A.K. Zvezdin and A.M. Kadomtseva, Sov. Sci. Rev. A 9 (1987) 117.
[5] C. Tabares-Munoz, J.-P. Rivera, A. Bezinges, A. Monnier and H. Schmid, Japan. J. Appl. Phys. 35 Suppl. 2 (1985) 1051.
[6] International tables for X-ray crystallography, Vol. 1 (Kynoch, Birmingham, 1965).
[7] A. Steiner, C. Tabares-Munoz and H. Schmid, Herbsttagung der SPG/SSP 60 (1987) p. 294.
[8] E. Asher, H. Rieder and H. Schmid, J. Appl. Phys. 37 (1966) 1404.